%% file: _Main.tex
\documentclass[pdflatex,sn-nature]{sn-jnl}

\usepackage{graphicx}%
\usepackage{multirow}%
\usepackage{amsmath,amssymb,amsfonts}%
\usepackage{amsthm}%
\usepackage[title]{appendix}%
\usepackage{xcolor}%
\usepackage{textcomp}%
\usepackage{manyfoot}%
\usepackage{booktabs}%
\usepackage{array}
\usepackage{algorithm}%
\usepackage{algorithmicx}%
\usepackage{algpseudocode}%
\usepackage{listings}%
\usepackage{xspace}

\usepackage{makecell}

\usepackage{subcaption}

\theoremstyle{thmstyleone}%

\theoremstyle{thmstyletwo}%

\theoremstyle{thmstylethree}%

\newcolumntype{L}[1]{>{\raggedright\let\newline\\\arraybackslash\hspace{0pt}}m{#1}}
\newcolumntype{C}[1]{>{\centering\let\newline\\\arraybackslash\hspace{0pt}}m{#1}}
\newcolumntype{R}[1]{>{\raggedleft\let\newline\\\arraybackslash\hspace{0pt}}m{#1}}

\newcommand{\ie}{\textit{i.e.,}\xspace}
\newcommand{\eg}{\textit{e.g.,}\xspace}

\newcommand{\att}{MNO-A\xspace}
\newcommand{\tmo}{MNO-B\xspace}
\newcommand{\vzw}{MNO-C\xspace}
\newcommand{\attandtmo}{MNO-A \& B\xspace}

\begin{document}

% \title[A Comprehensive Performance Comparison of Indoor Spectrum Sharing]{A Comprehensive Performance Comparison of Indoor Spectrum Sharing: Neutral-Host, Cellular Macro, and Wi-Fi}
% \title[Assessing Indoor Spectrum Sharing for Uplink]{Assessing Indoor Spectrum Sharing for Uplink: A Comparison Study of Deployment Testbeds}
\title{Indoor Neutral-Host Networks Over Shared Spectrum and Shared Infrastructure: A Comparison Study of Real-World Deployments}

\author*[1]{\fnm{Joshua Roy} \sur{Palathinkal}}\email{joshuaroy873@gmail.com}
\author[1]{\fnm{Muhammad Iqbal} \sur{Rochman}}\email{mrochman@nd.edu}
\author[2]{\fnm{Vanlin} \sur{Sathya}}\email{vanlin@celona.io}
\author[2]{\fnm{Mehmet} \sur{Yavuz}}\email{mehmet@celona.io}
\author[1]{\fnm{Monisha} \sur{Ghosh}}\email{mghosh3@nd.edu}

\affil[1]{\orgname{University of Notre Dame, Notre Dame, IN, USA}}
\affil[2]{\orgname{Celona, Inc., Campbell, CA, USA}}

\abstract{
% The increasing demand for high-capacity indoor connectivity is often at odds with the physical limitations of existing network deployments.
Indoor high-capacity connectivity is frequently constrained by significant building penetration loss and the inherent uplink power limitations of a typical outdoor macro-cell deployment. While Mobile Network Operators (MNOs) must optimize spectrum across low-band ($<$1~GHz) and mid-band (1--7~GHz) frequencies, uplink performance remains disproportionately degraded due to link budget asymmetry. Neutral-host (NH) networking provides a scalable alternative by transparently offloading MNO subscribers via spectrum sharing and shared infrastructure. We present a multi-site measurement study comparing Citizens Broadband Radio Service (CBRS)-enabled NH networks against public MNO 4G/5G macro deployments and Wi-Fi. Our results show: (i) significant building penetration loss with up to 15.5~dB in low-bands and 17.9~dB in mid-bands, resulting in a \(\sim\)10~dB RSRP deficit for MNO mid-bands compared to low-bands; (ii) NH networks provide a 30~dB higher median indoor RSRP with indoor NH normalized downlink throughput matches MNO outdoor performance, while its uplink performance exceeds MNO levels in both indoor and outdoor settings; (iii) NH proximity enables superior uplink efficiency, utilizing 64-QAM for 56\% of transmissions (versus $<$6\% for MNOs) and reducing median UE transmit power by 5~dB; (iv) MNOs rely on low-band spectrum for indoor uplink transmissions, while the NH deployment maintains high-performance mid-band connectivity; and (v) NH outperforms MNOs in end-to-end throughput but trails Wi-Fi in uplink throughput and latency due to packet routing overhead to the MNO core.
}


\maketitle

\input{_sections/introduction}
\input{_sections/related_works}
\input{_sections/methodology}
\input{_sections/results}
\input{_sections/conclusions}

\bibliography{_References}

\end{document}

%% file: _sections/introduction.tex
\section{Introduction}\label{intro}

Mobile network consumers expect indoor cellular performance to mirror the reliability of outdoor connectivity. Despite South Korea leading the world with 100\% 5G coverage and the highest density of 5G radio base station (gNB) (670 gNBs per 100,000 inhabitants)~\cite{EU5GObservatory2025}, user satisfaction for 5G remains lower than that of 4G, primarily due to inadequate indoor performance~\cite{Ericsson2021FiveWays5G} leading to a focus on improving indoor connectivity~\cite{Soth_Korea2025}.
In the U.S., mid-band spectrum (1--7~GHz) has become the preferred choice over low-band ($<$1~GHz) for delivering high-performance 5G, owing to its superior bandwidth availability. For instance, public Mobile Network Operators (MNOs) have utilized the 3.45--3.55~GHz and 3.7--3.98~GHz bands for commercial 5G deployments. However, consistent indoor connectivity is hindered by significant building penetration loss. Modern energy-efficient construction, particularly the use of low-emissivity (Low-E) glass, exacerbates this by substantially increasing RF attenuation at the mid-band frequencies critical for 5G.

Further, the performance of uplink is disproportionately degraded relative to an already attenuated downlink. This occurs because a gNB transmits at power levels hundreds of times higher than the maximum user equipment (UE) transmit power, making uplink the limiting link budget component. MNOs have attempted to mitigate these issues by deploying low-band ($<$1~GHz) to exploit better propagation characteristics and increase gNB site density~\cite{Ericsson2025MobilityReport}. However, these approaches merely alleviate symptoms rather than resolving the fundamental imbalance in indoor propagation and uplink constraints. 
Furthermore, operators may opt for solutions such as Distributed Antenna Systems (DAS) to improve indoor coverage. A DAS distributes cellular signals indoors but relies on an intricate network of passive cabling and specialized RF antennas, and are thus often cost-prohibitive, labor-intensive, and difficult to scale. Furthermore DAS deployments operate over MNO-owned licensed spectrum and each MNO has to manage their DAS deployment leading to increased costs.

\begin{figure}
    \centering
    \includegraphics[width=.7\linewidth]{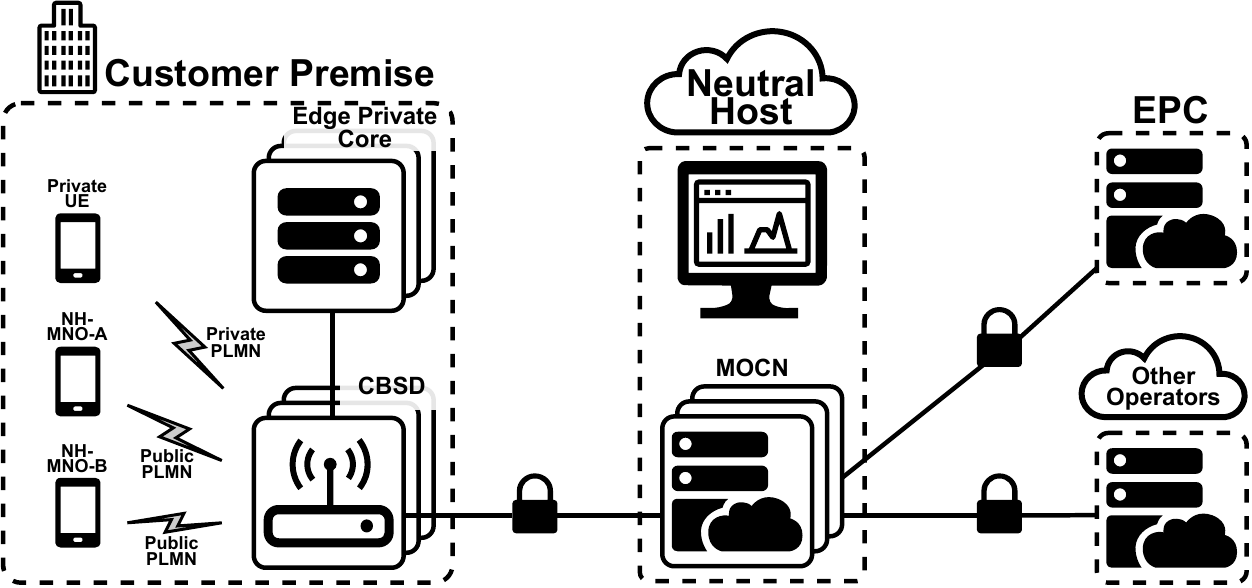}
    \caption{Neutral-host architecture~\cite{rochman2025neutral}.}
    \label{fig:nh_architecture}
    \vspace{-1em}
\end{figure}

Alternatively, neutral-host (NH) networking augments indoor coverage by leveraging shared-license spectrum, such as the 3.55~GHz Citizen Broadband Radio Service (CBRS) band in the U.S.~\cite{sathya2023nh}, along with a shared private network infrastructure. This approach enables private 4G/5G networks that can provide seamless indoor connectivity for public MNO subscribers.
Fig.~\ref{fig:nh_architecture} shows the architecture of a neutral-host network. Upon detecting a NH band, the UE retrieves a list of Public Land Mobile Network (PLMN) IDs from the broadcast network signature. This list represents a white-list of MNOs with active commercial agreements with the NH operator. If the subscriber's MNO is present, the device performs a handover to the neutral-host network without requiring user actions or authentication. Once connected to the CBRS base station (\ie CBSD/CBRS device), the subscriber's data packets are securely routed to the respective MNO core via a Multi-Operator Core Network (MOCN) gateway~\cite{3gpp23251}. This architecture establishes a secure tunnel, allowing the user to access data services authenticated by the SIM credentials associated with their existing service plan.
% Thus, rather than attempting to overcome severe penetration losses from macro gNB sites, this localized approach offers several key advantages:
% \begin{enumerate}
% \item \textbf{Infrastructure Sharing:} It enables multiple MNOs to utilize a single indoor infrastructure, eliminating the need for parallel deployments and significantly reducing operational overhead.
% \item \textbf{Seamless Mobility:} Handovers between NH nodes and the macro network are transparent, requiring no configuration changes, additional subscriptions, or user intervention.
% \item \textbf{Spectrum Coexistence:} By utilizing low-power indoor CBSD, the building's penetration loss effectively isolates indoor signals. This minimizes interference with outdoor incumbents and facilitates compatibility with shared-spectrum bands.
% \item \textbf{Capacity Offloading:} Indoor traffic is diverted to the NH network, freeing up valuable macro spectrum resources for outdoor MNO users.
% \end{enumerate}

The adoption of NH architecture offers a viable path to mitigate building penetration loss and uplink power constraints. This study contributes an empirical analysis of on-site measurements of co-located NH, 4G/5G MNO and 2.4/5~GHz Wi-Fi deployments at three distinct sites. This comparative data provides a foundational framework for future mid-band spectrum sharing studies, particularly for the 3.1--3.45 GHz and 7.125--8.4 GHz bands currently being studied by the National Telecommunications and Information Administration (NTIA) for shared use~\cite{ntia2023strategy}. While indoor NH networks can be deployed on any spectrum, overall spectrum utilization is maximized by reusing spectrum that has primarily outdoor incumbents, like in the CBRS band: incumbents can continue operation without interference while the spectrum can be shared by low-power indoor networks. Thus, the building loss, instead of being a liability, becomes an enabler of sharing spectrum.

Our contributions are as follows:

\begin{itemize}

% \noindent $\bullet$
\item 
\textbf{Comparison of building loss between three sites with distinct signal environments (\S\ref{sec_deployment},\S\ref{sec_building_loss}):} MNO macro deployments across all sites exhibit significant building penetration loss, reaching 15.5~dB in low-bands and 17.9~dB in mid-bands. Furthermore, median mid-band RSRP is consistently lower than low-band (\(\sim\)10.8~dB indoors and \(\sim\)10~dB outdoors). These results underscore the critical role of low-band spectrum in ensuring indoor coverage, even where mid-bands offer superior peak capacity.

% \noindent $\bullet$
\item 
\textbf{Comparison of coverage and physical layer performance (\S\ref{sec_building_loss},\S\ref{sec_phy_comp}):} The NH network exhibits significantly higher indoor signal strength, with a median RSRP of -89.9~dBm compared to -120.2~dBm for MNO macro deployments. While indoor MNO coverage is improved by a dedicated outdoor small-cell deployment, the neutral-host still provides a comparable normalized downlink throughput. In the uplink, the indoor NH solution outperforms MNO performance in both indoor and outdoor environments. 

% \noindent $\bullet$
\item 
\textbf{Focused analysis of uplink metrics (\S\ref{sec_uplink_metrics},\S\ref{sec_uplink_bands}):} Uplink performance data confirms a clear NH advantage, with 64-QAM utilized for 56\% of transmissions compared to less than 6\% for MNO macro networks. This efficiency is coupled with a 5~dB reduction in median UE TX power, driven by the shorter propagation paths of the NH deployment. The data also reveals a distinct shift in 5G band utilization by the MNOs---while mid-bands dominate outdoor traffic, their usage drops significantly indoors, highlighting that low-band spectrum remains the primary anchor for indoor uplink reliability.

% \noindent $\bullet$
\item 
\textbf{Comparison of end-to-end network performance (\S\ref{sec_app_reports}):} Our analysis reveals a performance trade-off where NH exceeds MNO indoor throughput but trails Wi-Fi in UL throughput and latency. We show that Wi-Fi median UL throughput is \(1.8\times\) higher than NH at matched 20~MHz bandwidth, an advantage that scales to \(10.7\times\) when Wi-Fi utilizes double the bandwidth of the NH network. This underscores the critical necessity for protocol and routing optimizations to achieve uplink and latency performance parity.

\end{itemize}

%% file: _sections/related_works.tex
\section{Related Work}\label{related}

There is a lack of extensive literature on NH network deployments, specifically regarding real-world coverage, coexistence, and performance.
For instance, \cite{bajracharya2022neutral, allawi2024cost} provide analyses of spectrum capacity and reuse based on deployment simulations. Other work  
\cite{fernandez2021validating, sathya2023nh, arendt2024distributed, arendt2022better} provides high-level performance analysis based on real-world deployment, but lacks detailed analyses of physical-layer (PHY) signal coverage and resource utilization.
On the other hand, our prior works~\cite{rochman2025neutral, palathinkal2025indoor} provides detailed breakdown of NH performance to generalize the benefits of NH infrastructure over the traditional MNO macro networks. These analyses are conducted based on measurements from different environments: \cite{rochman2025neutral} focuses on a hospital setting, while \cite{palathinkal2025indoor} analyzes a big-box retail location.
In both cases, the NH network operates on the CBRS band, which utilizes a spectrum-sharing model to coexist with outdoor incumbents, such as naval radar \cite{tusha2025comprehensive}. This framework creates a unique opportunity for low-power indoor private 4G/5G networks to operate without interfering with incumbents. Simultaneously, it provides robust indoor coverage that traditional outdoor MNO macro deployments struggle to achieve, while offering a vital avenue to offload heavy indoor traffic from the macro network. In this paper, we seek to extend the prior research by focusing specifically on uplink performance: it is becoming increasingly clear that many advanced 5G and future 6G applications will require improved uplink performance and many of these will be deployed indoors, such as factory and warehouse automation. 

%% file: _sections/methodology.tex
\section{Measurement Methodology \& Data Collection}\label{sec_methodology}

We utilized a Samsung S22+ smartphone capable of connecting to all relevant bands - 4G/5G cellular, Wi-Fi and CBRS neutral-host bands. QualiPoc and SigCap measurement tools were employed to obtain wireless Key Performance Indicator (KPI) metrics for comparative network evaluations. Passive KPIs (\eg signal strength metrics such as RSRP for LTE, SS-RSRP for NR, and RSSI for Wi-Fi) were continuously recorded by QualiPoc and SigCap, respectively. For active probe measurements, the QualiPoc tool executed a predefined test sequence comprising of: (i) a 5-second HTTP download from \textit{github.com} utilizing HTTP GET command; (ii) a 5-second HTTP upload to \textit{httpbin.org} utilizing HTTP PUT command; and (iii) ten ICMP echo requests (32-byte packets) to the \textit{Google DNS (8.8.8.8)} host. QualiPoc collects detailed LTE and NR PHY-layer measurements by directly probing the modem chipset through the \textit{Qualcomm Diagnostic Mode} interface~\cite{qualipoc}. In contrast, SigCap is limited to the information exposed by the Android API and was therefore used exclusively for Wi-Fi analysis~\cite{dogantusha2025spectrum}. Table~\ref{tab_qp_params} summarizes the QualiPoc and SigCap KPIs employed in this study.

\begin{table}
\centering
\caption{Summary of captured network performance parameters.}
\label{tab_qp_params}
\begin{tabular}{|p{0.28\linewidth}|p{0.67\linewidth}|}
\hline
\textbf{Parameter} & \textbf{Description} \\
\hline\hline

\multicolumn{2}{|c|}{\textbf{\textit{Qualipoc \& SigCap: General parameters}}} \\ \hline
Latitude, Longitude & UE's geographic coordinates \\ \hline

\multicolumn{2}{|c|}{\textbf{\textit{Qualipoc: Radio report parameters}}} \\ \hline
PCI & Physical Cell Identifier \\ \hline
DL/UL ARFCN & Absolute Radio Frequency Channel Number, i.e., center frequency \\ \hline
Bandwidth & Range of frequencies available for transmission [MHz] \\ \hline
RSRP & Reference Signal Received Power. For NR, RSRP is based on the Synchronization Signal (SS) block [dBm] \\ \hline
% SCS & Subcarrier spacing numerology; fixed at 15~kHz in LTE \\ \hline

\multicolumn{2}{|c|}{\textbf{\textit{Qualipoc: Performance \& power metrics}}} \\ \hline
% PDSCH/PUSCH throughput & Physical-layer throughput on the Physical Downlink/Uplink Shared Channel (downlink/uplink) [Mbps] \\ \hline
Normalized PDSCH/PUSCH throughput & PHY-layer throughput on the Physical Downlink/Uplink Shared Channel, normalized by allocated resource blocks, subcarrier spacing, and MIMO layers [bit/s/Hz/stream]~\cite{rochman2025comprehensive} \\ \hline
PDSCH/PUSCH modulation & Modulation scheme used on the Physical Shared Channel, \eg QPSK, 16-QAM, 64-QAM \\ \hline
% Path loss & Downlink path loss computed by UE for PUSCH power control [dB] \\ \hline
PUSCH TX power & UE's PUSCH uplink transmit power [dBm] \\ \hline
HTTP GET/PUT test status & Application-layer test outcome for HTTP downlink (GET) and uplink (PUT) transactions, indicating success or failure of end-to-end data transfer \\ \hline
IP DL/UL throughput & Network-layer throughput in downlink/uplink [Mbps] \\ \hline
Round-trip Time (RTT) & Measured RTT from ICMP ping  \\ \hline

% \multicolumn{2}{|c|}{\textbf{\textit{Qualipoc: PDSCH/PUSCH report parameters}}} \\ \hline
% \#RBs & Number of allocated resource blocks \\ \hline
% RBs per subframe/slot & Number of allocated Resource Blocks per subframe (LTE) or slot (NR) \\ \hline
% MCS & Modulation and coding scheme \\ \hline
% \#Layers & Number of MIMO layers used in DL transmissions \\ \hline
% BLER & Block error rate [\%] \\ \hline
% Path loss & Path loss calculated by the UE using reference signal transmit power and RSRP [dB] \\ \hline
% PUSCH TX power & UE uplink transmit power [dBm] \\ \hline

% \multicolumn{2}{|c|}{\textbf{\textit{Qualipoc: Calculated parameters}}} \\ \hline
% Normalized throughput & Throughput normalized by allocated resource blocks, subcarrier spacing, and MIMO layers [bit/s/Hz/stream] \cite{rochman2025comprehensive} \\ \hline

% \multicolumn{2}{|c|}{\textbf{\textit{Qualipoc: Application layer parameters}}} \\ \hline
% App. DL/UL throughput & Application-layer throughput in downlink and uplink [Mbps] \\ \hline
% TCP RTT & Round-trip time measured by TCP [ms] \\ \hline
% Ping RTT & Round-trip time measured by ICMP [ms] \\ \hline

\multicolumn{2}{|c|}{\textbf{\textit{SigCap: Wi-Fi parameters}}} \\ \hline
BSSID & Basic Service Set Identifier (unique identifier of a Wi-Fi AP) \\ \hline
Primary channel number & Primary channel associated with the BSSID [MHz] \\ \hline
RSSI & Received Signal Strength Indicator from the beacon signal [dBm] \\ \hline
% Channel utilization & Fraction of time the 20~MHz primary channel is busy [0,1] \\ \hline
% Wi-Fi station count & Number of Wi-Fi clients associated with the BSSID \\ \hline
TX power & Conducted transmit power of the BSSID [dBm] \\ \hline

\end{tabular}
\end{table}

\begin{figure}[t]
    \begin{subfigure}{.32\textwidth}
    \includegraphics[width=\linewidth, height=3.5cm]{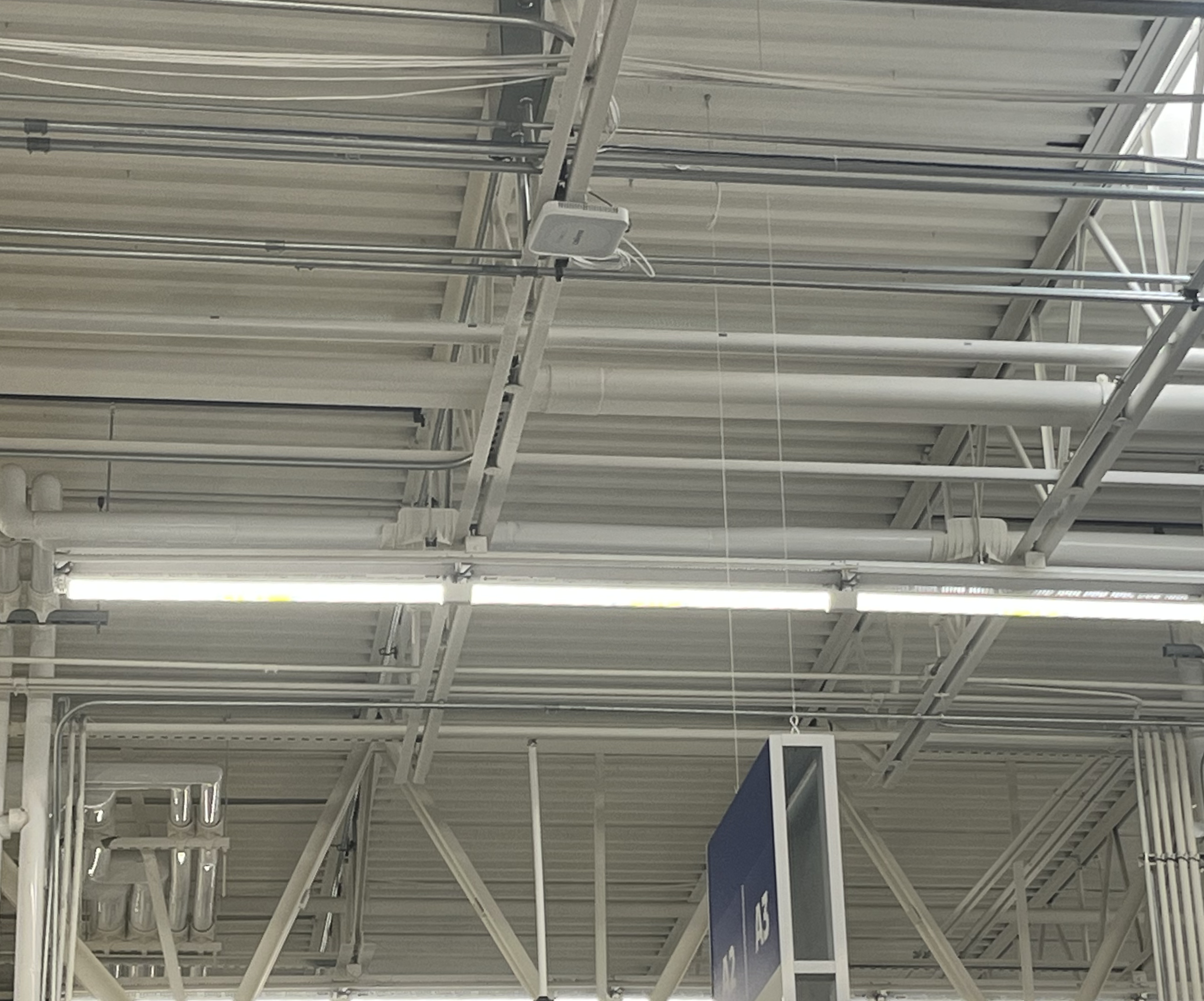}
    \caption{Site-A}
    \label{fig_environment_a}
    \end{subfigure}
    \hfill
    \begin{subfigure}{.32\linewidth}
    \includegraphics[width=\linewidth, height=3.5cm]{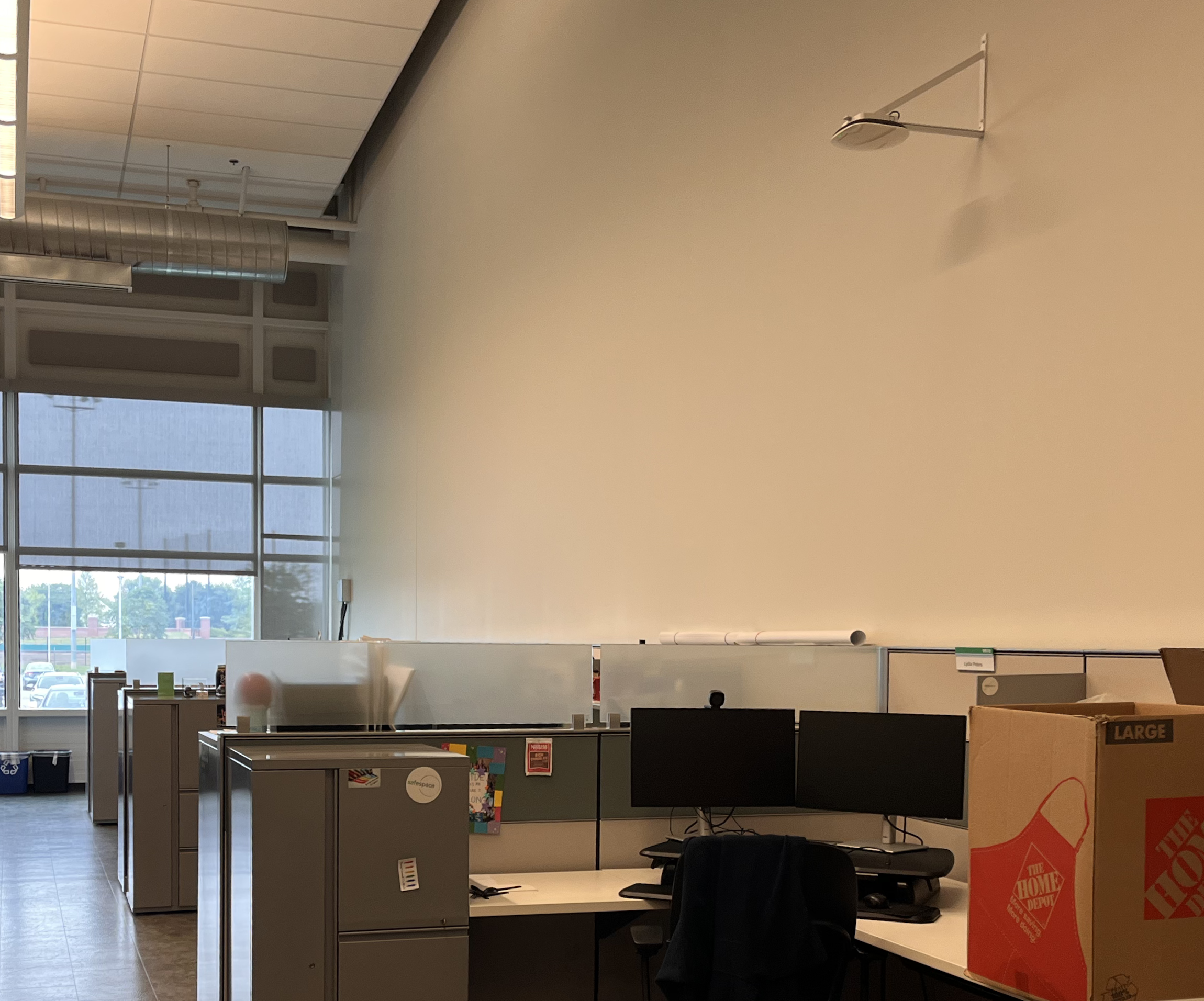}
    \caption{Site-B}
    \label{fig_environment_b}
    \end{subfigure}
    \hfill
    \begin{subfigure}{.32\textwidth}
    \includegraphics[width=\linewidth, height=3.5cm]{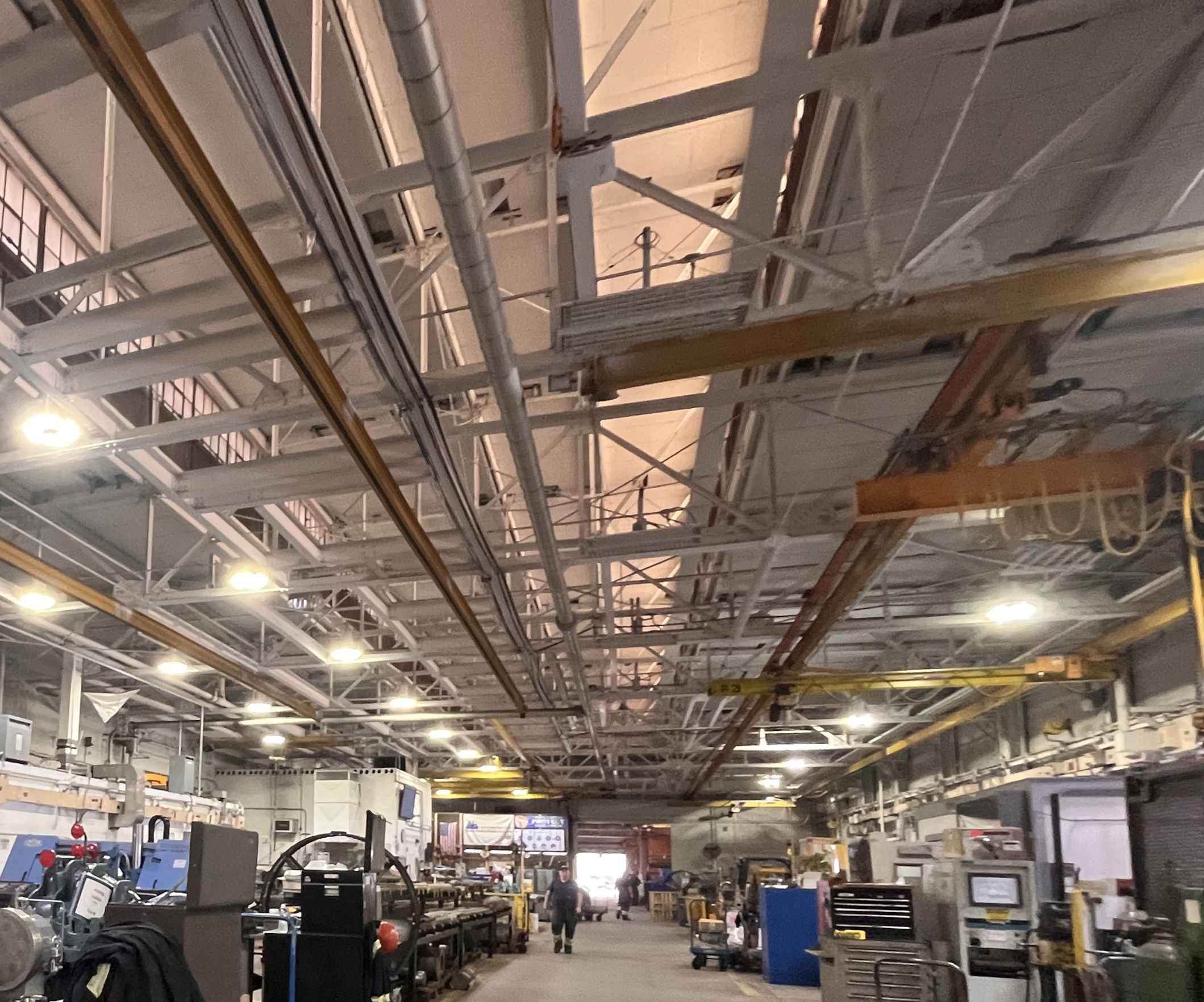}
    \caption{Site-C}
    \label{fig_environment_c}
    \end{subfigure}
    \caption{Overview of deployment sites.}
    \label{fig_environment}
    \vspace{-.5em}
\end{figure}

\begin{figure}
    \centering

    \begin{subfigure}[t]{0.32\textwidth}
        \centering
        \includegraphics[width=\linewidth]{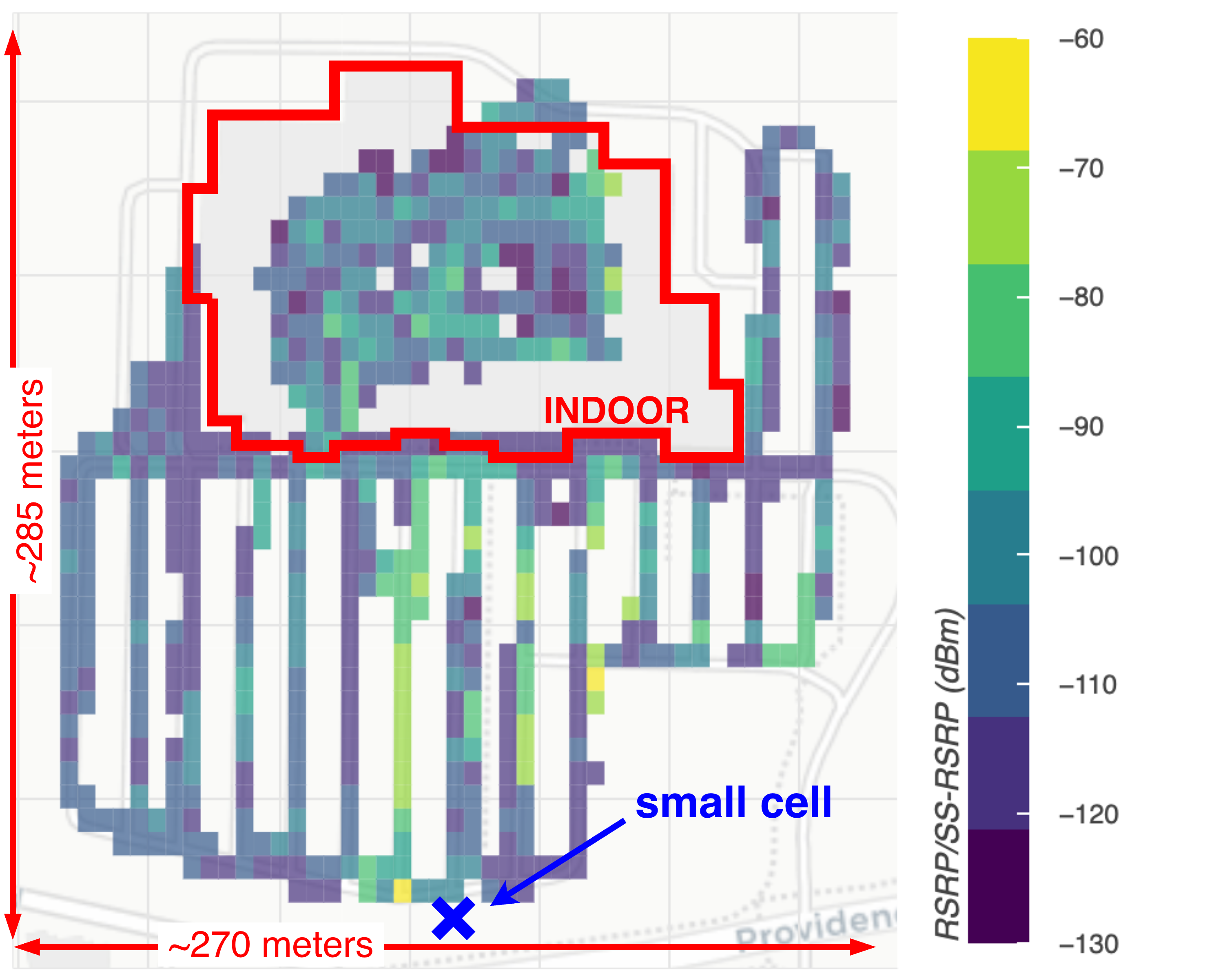}
        \caption{Site-A}
        \label{fig_5777AllCoverage}
    \end{subfigure}
    \hfill
    \begin{subfigure}[t]{0.32\textwidth}
        \centering
        \includegraphics[width=\linewidth]{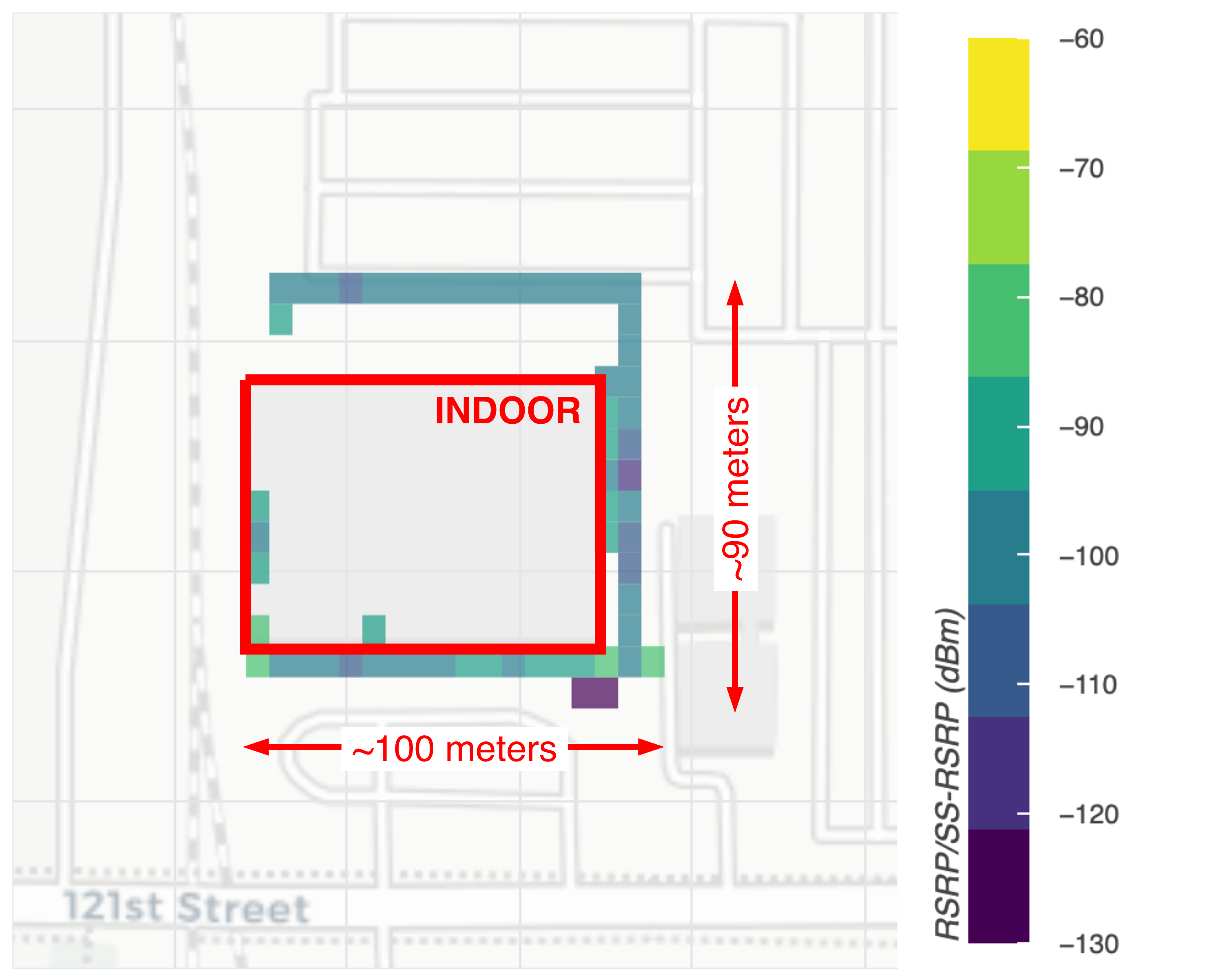}
        \caption{Site-B}
        \label{fig_warehouseAllCoverage}
    \end{subfigure}
    \hfill
    \begin{subfigure}[t]{0.32\textwidth}
        \centering
        \includegraphics[width=\linewidth]{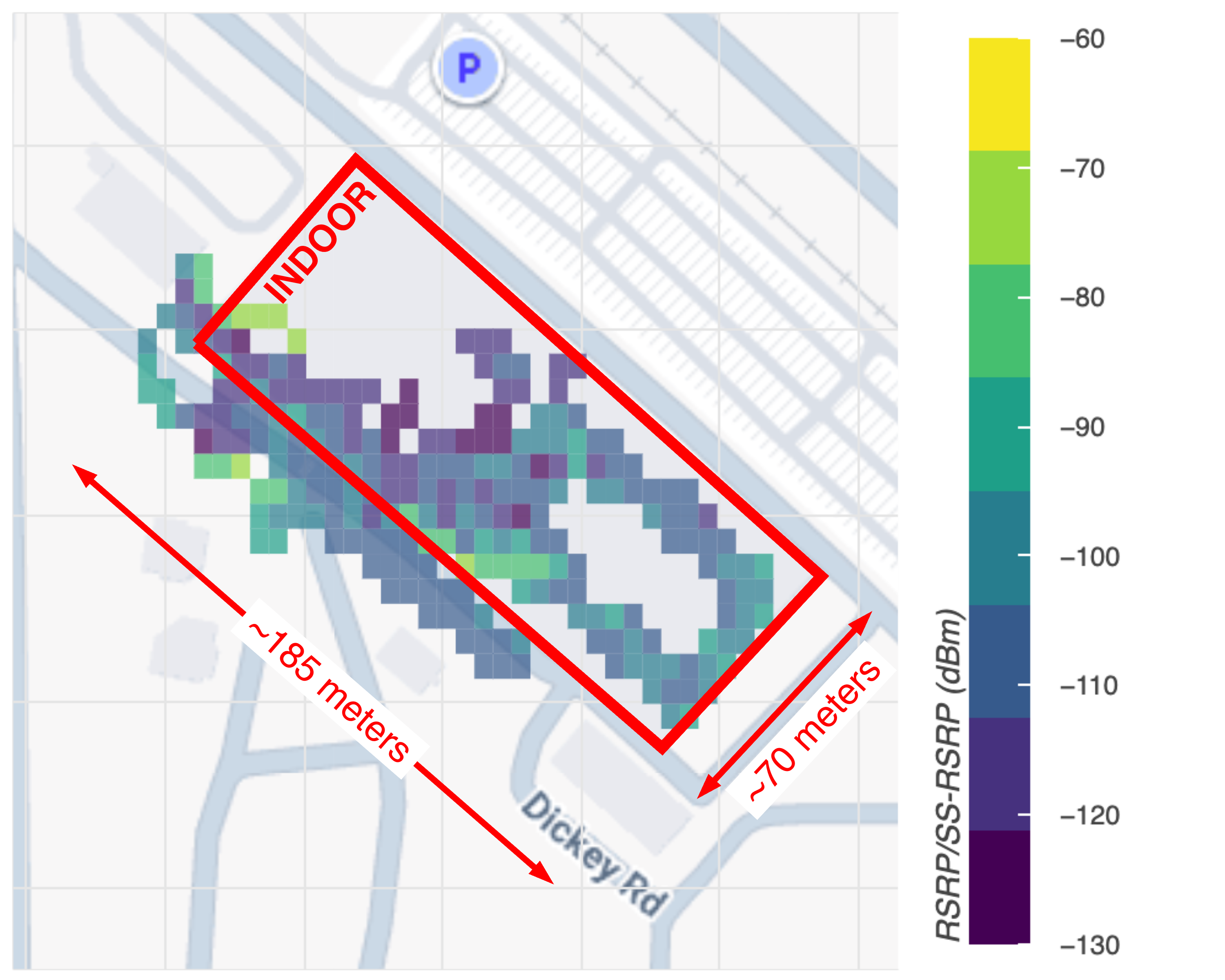}
        \caption{Site-C}
        \label{fig_adminAllCoverage}
    \end{subfigure}

    \caption{Cellular measurement footprint for Sites-A,B,C.}
    \label{fig_allCoverage}
\end{figure}

In this study, we measured data from two deployment locations: \textbf{a big-box retail store} and \textbf{a refinery complex}. Within the refinery complex, we further focus on two separate buildings, thus totaling three sites. Fig.~\ref{fig_environment} illustrates the three sites which represent distinct signal environments:
\begin{itemize}
    \item \textbf{Site-A} is a big-box retail store in the suburban area. This site is characterized by thick concrete walls and minimal windows. 
    \item \textbf{Site-B} is an office building within the refinery complex, characterized by large Low-emission (Low-e) glass windows and dry walls.
    \item \textbf{Site-C} is a warehouse annex in the refinery complex. It is characterized by metal structures and walls, as well as industrial equipments.
\end{itemize}
In all sites, both indoor CBSD and Wi-Fi Access Points (APs) are deployed facing down on the ceiling. 

The measurements were conducted by walking indoors and outdoors, in three separate campaigns between September and December of 2025, collecting a total of over 10 hours of data.
% capturing $17{,}400$~m$^2$ ($187{,}292$~ft$^2$) of a large big-box retail store indoors and the adjacent $28{,}400$~m$^2$ ($305{,}695$~ft$^2$) parking lot outdoors.
Three major cellular MNOs (labeled \att, \tmo, and \vzw) are available in all three sites, with only \attandtmo having commercial agreements to offload their subscribers to the neutral-host. In total, there are 254,728 datapoints captured over the QualiPoc radio reports, PDSCH (downlink) throughput, PUSCH (uplink) throughput, and application-layer (HTTP session and ping logs) reports, as well as 126,790 Wi-Fi beacon datapoints captured by SigCap.

\begin{table}
\centering
\captionsetup{font=small}
\caption{4G and 5G bands information.}
\vspace{-0.5em}
\label{tab_bands}
\begin{tabular}{C{.12\linewidth}C{.08\linewidth}C{.1\linewidth}C{.12\linewidth}C{.1\linewidth}C{.1\linewidth}}
\toprule
\textbf{Operator} & \textbf{Band} & \textbf{Duplex Mode} & \textbf{DL Band Freq. (MHz)} & \textbf{SCS (kHz)} & \textbf{DL BW (MHz)} \\

\midrule \midrule

\multicolumn{6}{c}{\textbf{NR bands}} \\ \midrule

\multirow{2}{*}{\att} & n5 & FDD & 850 & 15 & 10 \\
& n77 & TDD & 3700 & 30 & 40, 80 \\ \midrule

\multirow{2}{*}{\tmo} & n25 & FDD & 1900 & 15 & 10, 15, 20, 40 \\
& n41 & TDD & 2500 & 30 & 90, 100 \\
& n71 & FDD & 600 & 15 & 15, 20 \\ \midrule

\vzw & n77 & TDD & 3700 & 30 & 60, 100 \\ \midrule \midrule

\multicolumn{6}{c}{\textbf{LTE bands}} \\ \midrule

NH & b48 & TDD & 3500 & 15 & 20 \\ \midrule

\multirow{5}{*}{\att} & b2 & FDD & 1900 & 15 & 5, 10, 20\\
& b4 & FDD & 2100 & 15 & 20 \\
& b12 & FDD & 700 & 15 & 10 \\
& b14 & FDD & 700 & 15 & 5, 10 \\
& b30 & FDD & 2300 & 15 & 5, 10 \\ 
& b66 & FDD & 2100 & 15 & 5, 10 \\ \midrule

\multirow{5}{*}{\tmo} & b2 & FDD & 1900 & 15 & 5, 10 \\
& b4 & FDD & 2100 & 15 & 10, 20 \\
& b12 & FDD & 700 & 15 & 5, 10 \\
& b66 & FDD & 2100 & 15 & 20 \\
& b71 & FDD & 600 & 15 & 5 \\ \midrule

\multirow{6}{*}{\vzw} & b2 & FDD & 1900 & 15 & 15 \\
& b4 & FDD & 2100 & 15 & 20 \\
& b5 & FDD & 850 & 15 & 10 \\
& b13 & FDD & 700 & 15 & 10 \\
& b48 & TDD & 3500 & 15 & 10, 20 \\
& b66 & FDD & 2100 & 15 & 20 \\ \bottomrule

\end{tabular}
\end{table}

\section{Deployment Overview}\label{sec_deployment}

Fig.~\ref{fig_allCoverage} illustrates the cellular measurement footprints for Sites-A,B,C, aggregating measurements across all deployments (macro and neutral-host), operators (MNO-A,B,C), and radio access technologies (LTE, NR). The maps depict the average signal strength, computed using RSRP for LTE and SS-RSRP for NR, and spatially averaged over $7.5$~m$\times$ $7.5$~m square bins.

While Site-A provided comprehensive indoor and outdoor coverage, Sites-B,C exhibited spatial constraints. Site-B's outdoor measurements were restricted to the building's immediate perimeter, while Site-C were restricted to the immediate west of the building. Further, Site-B's indoor data suffered from significant GPS attrition: approximately 69,000 data points across varying timestamps reported identical coordinates. This indicates a GPS failure that limited the usable indoor measurement region. Despite this lack of spatial data, these data points are retained in our analysis as the study focuses on performance metrics rather than spatial correlations.

\begin{table}
\centering
\captionsetup{font=small}
\caption{MNO Macro deployment summary (bands in bold indicate mid-bands).}
\label{tab_mno_deployment}
\begin{tabular}{llcccc} % Removed vertical bars for a cleaner look
\toprule
\textbf{Operator} & \textbf{IN/OUT} & \multicolumn{2}{c}{\textbf{Bands}} & \multicolumn{2}{c}{\textbf{\# PCI}} \\ 
\cmidrule(lr){3-4} \cmidrule(lr){5-6} % Use cmidrule instead of cline
& & \textbf{LTE} & \textbf{NR} & \textbf{LTE} & \textbf{NR} \\ 
\midrule \midrule

\multicolumn{6}{c}{\textbf{Site-A}} \\ \midrule
\multirow{2}{*}{\att} & IN  & \textbf{b2}, b12, b14, \textbf{b30}, \textbf{b66} & n5 & 65 & 1 \\
                       & OUT & \textbf{b2}, \textbf{b4}, b12, b14, \textbf{b30}, \textbf{b66} & n5, \textbf{n77} & 93 & 5 \\ \addlinespace
\multirow{2}{*}{\tmo} & IN  & \textbf{b2}, \textbf{b4}, b12, b71 & \textbf{n25}, \textbf{n41}, n71 & 67 & 3 \\
                       & OUT & \textbf{b2}, \textbf{b4}, b12, \textbf{b66} & \textbf{n25}, \textbf{n41}, n71 & 39 & 4 \\ \addlinespace
\multirow{2}{*}{\vzw} & IN  & \textbf{b2}, \textbf{b4}, b5, b13, \textbf{b66} & -- & 14 & -- \\
                       & OUT & \textbf{b2}, \textbf{b4}, b5, b13, \textbf{b48}, \textbf{b66} & \textbf{n77} & 51 & 1 \\ \midrule \midrule

\multicolumn{6}{c}{\textbf{Site-B}} \\ \midrule
\multirow{2}{*}{\att} & IN  & \textbf{b2}, \textbf{b4}, b12, \textbf{b30}, \textbf{b66} & n5, \textbf{n77} & 16 & 4 \\
                       & OUT & \textbf{b2}, \textbf{b4}, b12, b14, \textbf{b30}, \textbf{b66} & \textbf{n77} & 43 & 1 \\ \addlinespace
\multirow{2}{*}{\tmo} & IN  & \textbf{b2}, \textbf{b4}, b12, \textbf{b66}, b71 & \textbf{n41}, n71 & 20 & 4 \\
                       & OUT & \textbf{b2}, \textbf{b66} & \textbf{n25}, \textbf{n41}, n71 & 13 & 4 \\ \addlinespace
\multirow{2}{*}{\vzw} & IN  & \textbf{b4}, b13 & -- & 14 & -- \\
                       & OUT & \textbf{b5}, b13, \textbf{b48}, \textbf{b66} & \textbf{n77} & 20 & 3 \\ \midrule \midrule

\multicolumn{6}{c}{\textbf{Site-C}} \\ \midrule
\multirow{2}{*}{\att} & IN  & \textbf{b2}, \textbf{b4}, b12, b14, \textbf{b30}, \textbf{b66} & \textbf{n77} & 43 & 2 \\
                       & OUT & \textbf{b2}, b12, b14, \textbf{b30}, \textbf{b66} & -- & 58 & -- \\ \addlinespace
\multirow{2}{*}{\tmo} & IN  & \textbf{b2}, \textbf{b4}, b12, b71 & \textbf{n41}, n71 & 27 & 5 \\
                       & OUT & -- & \textbf{n25}, \textbf{n41}, n71 & -- & 3 \\ \addlinespace
\multirow{2}{*}{\vzw} & IN  & \textbf{b4}, b5, b13, \textbf{b48}, \textbf{b66} & \textbf{n77} & 50 & 1 \\
                       & OUT & \textbf{b4}, b5, b13, \textbf{b48}, \textbf{b66}& \textbf{n77} & 41 & 3 \\ 
\bottomrule
\end{tabular}
\end{table}

Table~\ref{tab_bands} highlights the distinct physical layer configuration across operators. For all operators, 4G bands (denoted with the prefix~\textit{`b'}) typically used FDD% and 15~kHz SCS
, except for the TDD-based b48 (\ie the CBRS band). Most 5G mid-bands utilized TDD and 30~kHz SCS, though n25 followed the low-band convention of FDD and 15~kHz SCS. Channel bandwidths for 4G and 5G low-bands typically ranged between 5 and 20~MHz, whereas 5G mid-bands (e.g., n41 and n77) reached up to 100~MHz. While 5G Non-Standalone (NSA) mode was the standard across all MNOs, \tmo additionally utilized Standalone (SA) mode. Notably, the CBRS band (designated as band b48 per 3GPP standards) were not only utilized by the private NH network, but also by \vzw in a macro-cell configuration.

\subsection{MNO Macro}\label{sec_deployment_mno_macro}

Site-specific variations in band configurations for both indoor and outdoor environments are summarized in Table~\ref{tab_mno_deployment}. Based on the count of unique Physical Cell Identifiers (PCIs), we observe a significantly lower NR deployment density compared to LTE. Notably, mid-band frequencies (indicated in bold) exhibit reduced indoor availability, with fewer unique PCIs detected indoors than outdoors. Regarding network architecture, although \tmo utilizes both 5G SA and NSA modes, SA mode was observed exclusively in outdoor measurements. Furthermore, \vzw’s CBRS (b48) deployments were primarily restricted to outdoor regions. These observations are consistent with the expected effects of building penetration loss on mid-band and high-frequency signal propagation.

\begin{table}
\centering
\captionsetup{font=small}
\caption{Neutral-host deployment parameters across measurement sites.}
\label{tab_nh_deployment}
\begin{tabular}{|p{4cm}|c|c|c|}
\hline
\rule{0pt}{3ex} \textbf{Parameter} & \textbf{Site-A} & \textbf{Site-B} & \textbf{Site-C} \\ \hline \hline

\textbf{\#CBSDs / \#PCIs} & 5 / 10 & 3 / 6 & 4 / 8 \\ \hline
\textbf{Antenna type} & \multicolumn{3}{c|}{Omnidirectional} \\ \hline
\textbf{4G TDD config} & \multicolumn{3}{c|}{Config \#1: 4 DL \& 4 UL subframes} \\ \hline
\textbf{CBSD transmit power} & 23, 24~dBm & \multicolumn{2}{c|}{24~dBm} \\ \hline
\textbf{Antenna gain} & \multicolumn{3}{c|}{6~dBi} \\ \hline
\textbf{Center frequency} &
\begin{tabular}[c]{@{}c@{}}
\{3590, 3610, 3630,\\
3650, 3670, 3690\}~MHz
\end{tabular}
& \multicolumn{2}{c|}{\{3670, 3690\}~MHz} \\ \hline
\textbf{Configured bandwidth} & \multicolumn{3}{c|}{20~MHz} \\ \hline
\textbf{Channel aggregation} & \multicolumn{3}{c|}{2 channels in downlink (40~MHz), intra-CBSD} \\ \hline

\end{tabular}
\end{table}

\subsection{CBRS Neutral-host}\label{sec_deployment_nh}

The neutral-host infrastructure at all three sites utilizes identical indoor CBSD models provided by Celona, Inc. Each unit is configured with two PCIs at a 20 MHz channel bandwidth, enabling intra-CBSD carrier aggregation (CA) for a total bandwidth of 40 MHz; notably, inter-CBSD CA is not supported. These CBSDs feature a maximum possible transmit power of 24 dBm and utilize internal omnidirectional antennas with a 6~dBi gain.
However, actual configurations for each CBSDs are subject to Spectrum Access System (SAS) grants. For instance, 3 out of 10 PCIs at Site-A operated at 23 dBm transmit (TX) power with access to the entire CBRS band. In contrast, all PCIs at Sites-B,C were granted 24~dBm but limited to the upper portion of the band (3660--3700~MHz). A comprehensive summary of the deployment configuration is provided in Table~\ref{tab_nh_deployment}.

\subsection{Enterprise Wi-Fi}\label{sec_deployment_wifi}

Table~\ref{tab_wifi_deployment} summarizes the enterprise Wi-Fi deployments across the three test sites. While all sites conventionally used the 2.4 and 5 GHz bands with channel widths up to 40 MHz, the deployments vary significantly in terms of TX power and supported standards (\eg 802.11n, 11ac, and 11ax). Notably, Site-A features the lowest Wi-Fi AP density (identified by the number of Basic Service Set Identifiers/BSSIDs), despite having the largest indoor area (14,000~m$^2$) compared to Sites-B and C (5,000~m$^2$ and 13,000~m$^2$, respectively). Additionally, Site-C utilizes two distinct deployments from the same provider: the first deployment identified by the 802.11n support in 2.4 GHz and 802.11ac support in 5 GHz, while the second is identified by 802.11ax support in both 2.4 and 5 GHz. However, the network is heavily dominated by the latter deployments, which accounts for approximately 99\% of the collected data across both bands.

\begin{table}
\centering
\captionsetup{font=small}
\caption{Wi-Fi deployment parameters across measurement sites.}
\label{tab_wifi_deployment}
\begin{tabular}{|c|c|c|c|c|c|}
\hline
\rule{0pt}{3ex}
\textbf{Site} &
\textbf{Band} &
\textbf{\# BSSIDs} &
\textbf{Bandwidth} &
\textbf{TX power} &
\textbf{Standard} \\ \hline \hline

\multirow{2}{*}{Site-A}
 & 2.4~GHz & 13 & 20~MHz & 14~dBm & 802.11ax \\
 & 5~GHz   & 54 & 20~MHz & 10--16~dBm & 802.11ax \\ \hline

\multirow{2}{*}{Site-B}
 & 2.4~GHz & 73 & 20~MHz & 10--11~dBm & 802.11n \\
 & 5~GHz   & 171 & 40~MHz & 14--16~dBm & 802.11ac \\ \hline

\multirow{2}{*}{Site-C}
 & 2.4~GHz & 118 & 20~MHz & 3--27~dBm & 802.11n/ax \\
 & 5~GHz   & 112 & 40~MHz & 6--26~dBm & 802.11ac/ax \\ \hline

\end{tabular}
\end{table}

%% file: _sections/results.tex
\section{Results}\label{results}

\subsection{Cellular Coverage and Building Loss}\label{sec_building_loss}

\begin{figure}
    \centering
    \includegraphics[width=0.5\linewidth]{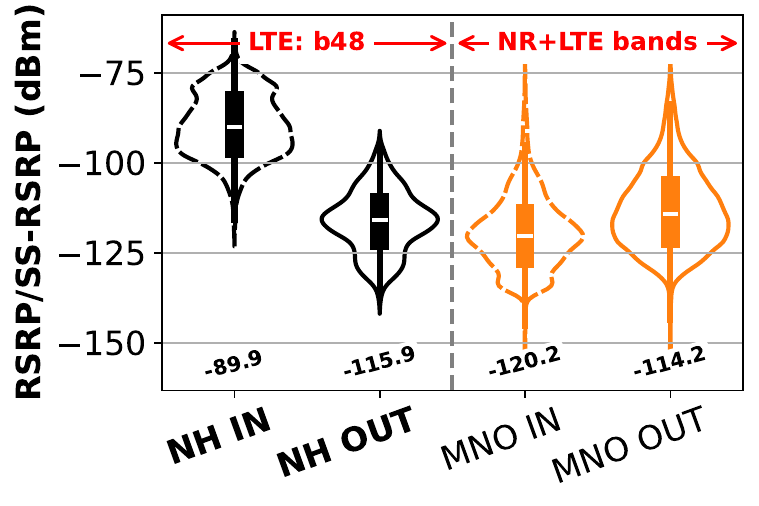}
    \caption{RSRP/SS-RSRP comparison over all sites.}
    \label{fig_rsrpInOutNhMno}
\end{figure}

Fig.~\ref{fig_rsrpInOutNhMno} illustrates the RSRP/SS-RSRP distributions for neutral-host and MNO macro networks across all evaluated operators, bands, and sites. As expected for an indoor-deployed solution, the NH network exhibits significantly higher indoor signal strength, with a median RSRP of -89.9~dBm compared to \mbox{-120.2}~dBm for MNO macros. For MNO macro deployments, the median indoor-to-outdoor signal strength difference---the measured building loss \(\Delta\)---is relatively small at approximately 6~dB. This low figure is attributed to the aggregation of various low- and mid-band frequencies in the MNO dataset. In contrast, the neutral-host deployment exhibits a substantially larger median indoor-to-outdoor difference of 26~dB---a predictable higher loss due to higher frequency.

To provide detailed analysis between low- and mid-band MNO macro, Table~\ref{tab_inOutSiteCoverage} provides the summary of building loss \(\Delta\) as well as the median, 25th, and 75th percentile values (P25/P75) of RSRP for each site. We observe higher \(\Delta\) at mid-band frequencies than at low-band for Sites-A,B, as expected from the increased attenuation characteristics at higher frequencies. In contrast, this trend is not observed for Site-C, likely due to the limited outdoor measurement footprint at that site (illustrated in Fig.~\ref{fig_adminAllCoverage}), which restricts the ability to capture representative outdoor coverage conditions.

Frequency-dependent propagation characteristics create a measurable RSRP gap between bands. As detailed in Table~\ref{tab_inOutSiteCoverage}, the median RSRP for mid-band is approximately 10.8~dB lower than low-band in indoor environments and 10~dB lower in outdoor settings. From a network management perspective, this implies that a UE undergoing an inter-frequency transition from low- to mid-band will see its RSRP reduced by roughly 10~dB. Such a substantial disparity highlights why low-band spectrum remains critical for indoor coverage, even as MNOs prioritize mid-bands for high-throughput 5G services.
% Across Sites-A,B,C, a pronounced additional attenuation is observed when transitioning from low-band to mid-band operation. From the median values reported in Table~\ref{tab_inOutSiteCoverage}, the inferred signal strength degradation between low-band and mid-band frequencies reaches up to 10.8~dB for indoor measurements and approximately 10~dB for outdoor measurements.

\begin{table}
\centering
\captionsetup{font=small}
\caption{Comparison of RSRP/SS-RSRP across all MNOs and locations.}
\label{tab_inOutSiteCoverage}
\begin{tabular}{L{1.8cm}L{1.4cm}C{1.4cm}ccc}
\toprule
\textbf{Site} & \textbf{Band} & \textbf{IN/OUT} & \multicolumn{3}{c}{\textbf{RSRP / SS-RSRP (dBm)}} \\
\cmidrule(lr){4-6}
& & & \textbf{Median} & \textbf{P25 / P75} & \textbf{\(\Delta\)} \\
\midrule\midrule

\multirow{4}{*}{\textbf{Site-A}} 
& \multirow{2}{*}{low-band} & IN  & -121.4 & -126.2 / -116.5 & \multirow{2}{*}{9.9} \\
&  & OUT & -111.5 & -117.1 / -106.2 & \\ \addlinespace
& \multirow{2}{*}{mid-band} & IN  & -132.2 & -134.6 / -125.6 & \multirow{2}{*}{10.7} \\
&  & OUT & -121.5 & -125.9 / -116.9 & \\
\midrule\midrule

\multirow{4}{*}{\shortstack[l]{\textbf{Small-cell}\\\textit{(Site-A)}}}
& \multirow{2}{*}{low-band} & IN  & -86.9 & -90.4 / -82.6 & \multirow{2}{*}{9.1} \\
&  & OUT & -77.8 & -84.8 / -74.0 & \\ \addlinespace
& \multirow{2}{*}{mid-band} & IN  & -93.9 & -97.7 / -90.1 & \multirow{2}{*}{9.9} \\
&  & OUT & -84.0 & -90.9 / -76.3 & \\
\midrule\midrule

\multirow{4}{*}{\textbf{Site-B}} 
& \multirow{2}{*}{low-band} & IN  & -112.4 & -115.8 / -108.5 & \multirow{2}{*}{15.5} \\
&  & OUT & -96.9  & -100.7 / -86.7  & \\ \addlinespace
& \multirow{2}{*}{mid-band} & IN  & -117.6 & -122.6 / -111.1 & \multirow{2}{*}{17.9} \\
&  & OUT & -99.7  & -108.3 / -91.9  & \\
\midrule\midrule

\multirow{4}{*}{\textbf{Site-C}} 
& \multirow{2}{*}{low-band} & IN  & -112.4 & -117.1 / -105.3 & \multirow{2}{*}{11.1} \\
&  & OUT & -101.3 & -104.2 / -99.1  & \\ \addlinespace
& \multirow{2}{*}{mid-band} & IN  & -116.8 & -121.9 / -111.2 & \multirow{2}{*}{9.2} \\
&  & OUT & -107.6 & -110.8 / -103.6 & \\
\bottomrule
\end{tabular}
\end{table}

\subsubsection{Indoor Coverage Enhancement via Dedicated Small-Cells}

To bolster indoor performance at Site-A, \att deployed three outdoor-mounted LTE small-cells operating on bands b2 (PCI-488), b12 (PCI-336), and b66 (PCI-101). These units operate with a 10 MHz channel bandwidth and a 4\(\times\)4 MIMO configuration. The small-cell antennas are oriented northward toward the retail facility, with their specific locations detailed in Fig.~\ref{fig_5777AllCoverage}. While these small-cells were deactivated during the aggregate MNO coverage analysis presented in the previous section, they were enabled for this portion of the study to isolate and evaluate their specific impact on the indoor signal environment.

The performance of the small-cell deployment is summarized in Table~\ref{tab_inOutSiteCoverage}. While the measured building loss \(\Delta\) remains comparable to macro-only deployments---as the signals still originate from an outdoor source---the localized presence of outdoor small-cells substantially elevates the absolute indoor signal strength. Specifically, the median indoor RSRP improves from -121.4~dBm to -86.9~dBm in low-band operation and from -132.2~dBm to -93.9~dBm in mid-band. Consistent with our earlier findings, the median RSRP of mid-band frequencies are lower than low-band of approximately 7~dB indoors and 6.2~dB outdoors.

\subsubsection{PCI-Specific Analysis}

In addition to aggregate signal strength analysis, measurements were also examined on a per-PCI basis to enable a more controlled comparison by focusing on signals from a single base-station source. This approach helps reduce variability associated with mixed cell origins and offers clearer insight into frequency-dependent indoor attenuation.

\begin{figure}
    \centering
    \includegraphics[width=0.5\linewidth]{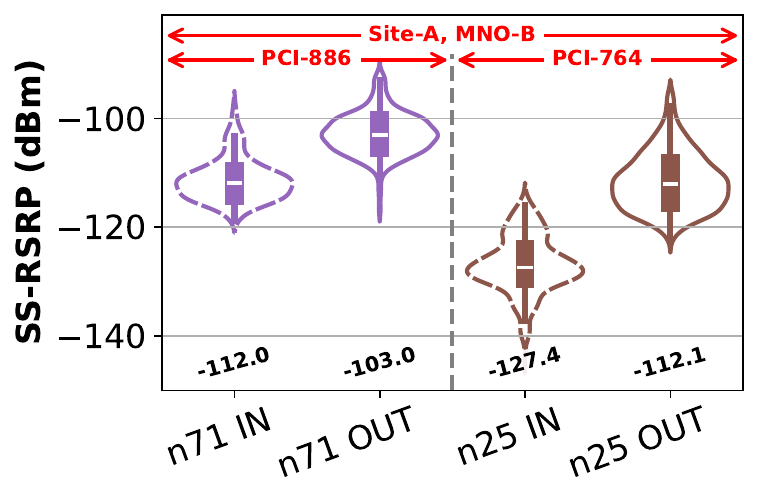}
    \caption{Indoor-outdoor RSRP comparison for MNO-B NR low- and mid-band PCIs at Site-A.}
    \label{fig_rsrpInOutSiteAMnoB}
\end{figure}

We first consider NR operation at Site-A, where MNO-B offers measurements spanning both low- and mid-band NR carriers indoors and outdoors. Specifically, PCI-886 (n71, 600~MHz) representing low-band and PCI-764 (n25, 1900MHz) representing mid-band were selected. The corresponding SS-RSRP distributions are shown in Fig.\ref{fig_rsrpInOutSiteAMnoB}. The median building loss increases from approximately 9~dB at low-band to 15.3~dB at mid-band, indicating substantially higher indoor attenuation at higher frequencies. Although this NR comparison is performed per PCI, the two bands are transmitted from different cells.

\begin{figure}
    \centering
    \includegraphics[width=0.6\linewidth]{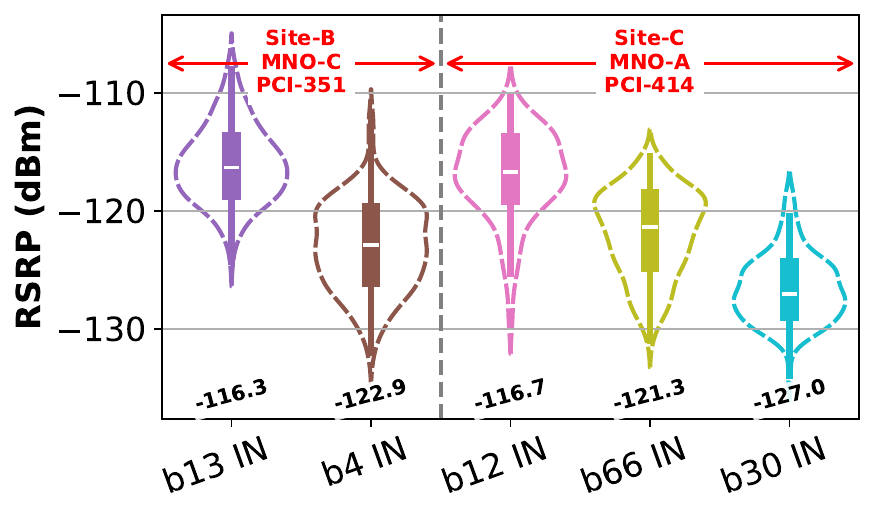}
    \caption{RSRP comparison across LTE bands sharing common PCI at Sites-B,C.}
    \label{fig_rsrpInSiteBC}
\end{figure}

To further isolate frequency effects, we analyze LTE deployments where multiple bands are transmitted from the same PCI. This configuration enables a direct comparison of indoor degradation across frequencies originating from an identical source. Two representative cases were selected: PCI-351 at Site-B (MNO-C), operating b13 (700~MHz) and b4 (1700~MHz), and a PCI at Site-C (MNO-B), operating b12 (700~MHz), b66 (1700~MHz), and b30 (2300~MHz). These PCIs provide sufficient indoor measurement density for comparison. As shown in Fig.\ref{fig_rsrpInSiteBC}, indoor signal degradation consistently increases with operating frequency, reinforcing the frequency-dependent trends observed in earlier analyses.

\subsection{Cellular PHY-Layer Performance Comparison}\label{sec_phy_comp}

\begin{figure}
    \centering

    \begin{subfigure}[t]{0.48\textwidth}
        \centering
        \includegraphics[width=\linewidth]{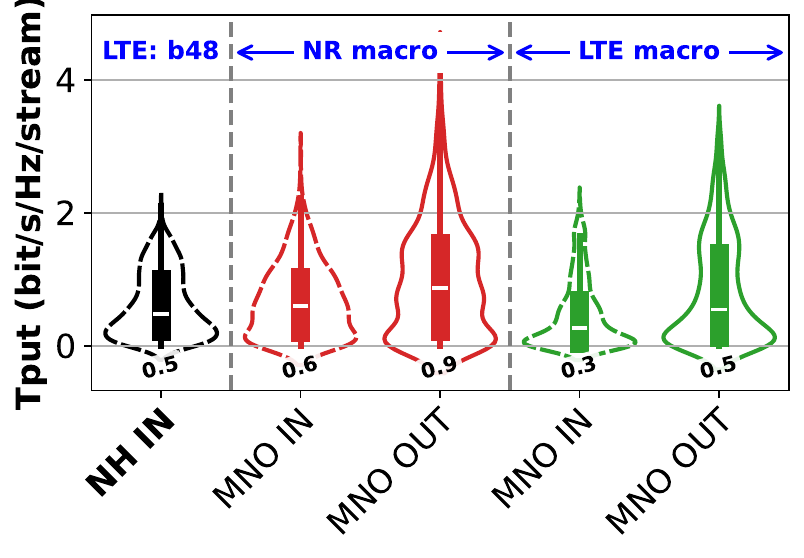}
        \caption{Normalized PDSCH throughput}
        \label{fig_phyDlNormTputMnoNh}
    \end{subfigure}
    \hfill
    \begin{subfigure}[t]{0.48\textwidth}
        \centering
        \includegraphics[width=\linewidth]{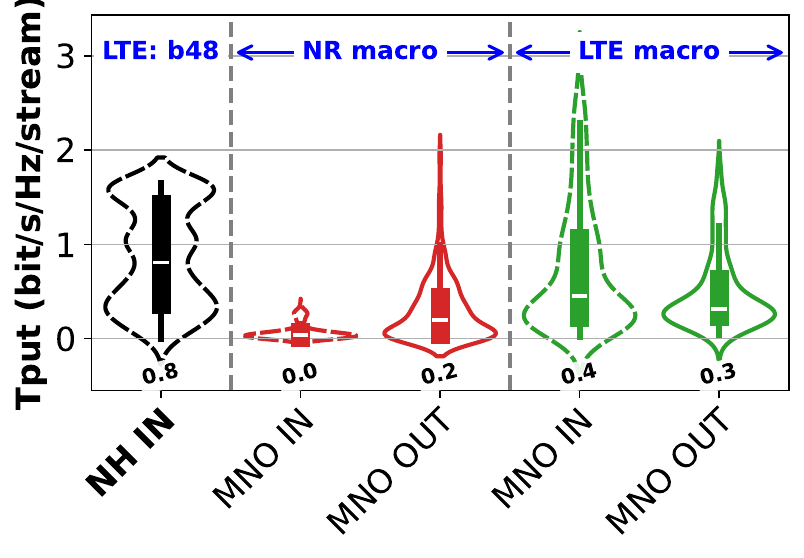}
        \caption{Normalized PUSCH throughput}
        \label{fig_phyUlNormTputMnoNh}
    \end{subfigure}

    \caption{Normalized throughput comparison over all sites}
    \label{fig_phyNormTputMnoNh}
\end{figure}

Fig.~\ref{fig_phyNormTputMnoNh} presents the normalized PHY-layer throughput performance across all operators and sites. This metric represents spectral efficiency thus enabling a fair comparison across heterogeneous deployments~\cite{rochman2025comprehensive}, \ie
\begin{equation*}
T_{\text{norm}} = \frac{T_{\text{bps}}}{N_{\text{RB}} \cdot SCS_{\text{Hz}} \cdot 12 \cdot N_{\text{layer}}},
\label{eq_norm_tput}
\end{equation*}
where $T_{\text{bps}}$ is the measured throughput in bits/second, $N_{\text{RB}}$ is the average number of RBs allocated to the UE per subframe(LTE)/slot(NR), $SCS_{\text{Hz}}$ is the subcarrier spacing in Hz, and $N_{\text{layer}}$ is the number of MIMO layers used. The term $N_{\text{RB}} \cdot SCS_{\text{Hz}} \cdot 12$ represents the utilized bandwidth, accounting for the 12 subcarriers configured per RB. The results aggregate performance across both low- and mid-band spectrum and are calculated on a per-carrier basis. In Fig.~\ref{fig_phyNormTputMnoNh}, MNO macro performance for both LTE and NR is presented alongside the neutral-host results for a comparative analysis.

In the downlink, Fig.~\ref{fig_phyDlNormTputMnoNh} indicates no statistically significant difference between indoor neutral-host performance and indoor MNO NR or outdoor MNO LTE macro deployments. In effect, the neutral-host system delivers downlink PHY-layer performance comparable to that of an outdoor LTE macro network. Similar observations have been reported at a different measurement site, as documented in~\cite{palathinkal2025indoor}.
The uplink advantage is evident in Fig.~\ref{fig_phyUlNormTputMnoNh}, where the NH network delivers substantially higher normalized throughput indoors compared to MNO macro deployments.

\subsubsection{Comparison of Uplink Metrics}\label{sec_uplink_metrics}

\begin{figure}
    \centering
    \includegraphics[width=0.6\linewidth]{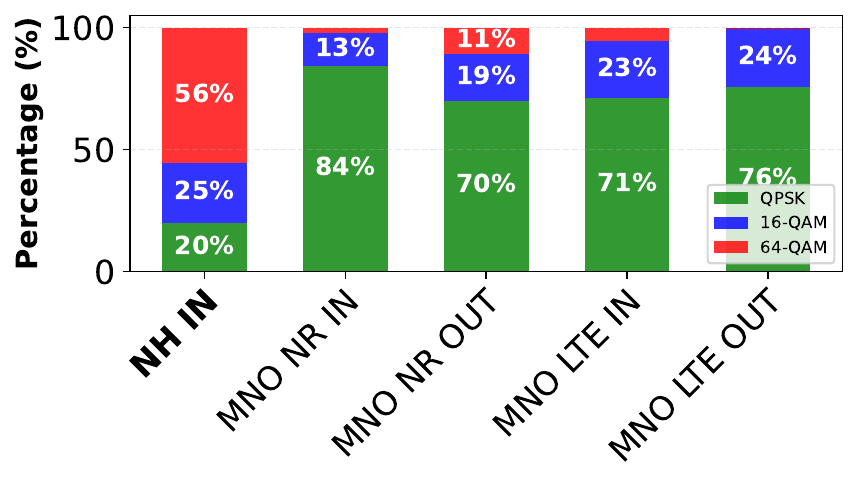}
    \caption{Uplink modulation usage distribution.}
    \label{fig_phyUlModMnoNh}
\end{figure}

The uplink advantage of neutral-host is further supported by the modulation statistics in Fig.~\ref{fig_phyUlModMnoNh}, which demonstrate a predominant use of higher-order uplink modulation for the neutral-host across Sites-A,B,C. Notably, 64-QAM was employed for approximately 56\% of the NH uplink transmission time. Conversely, indoor MNO macro rarely utilized high-order modulations, with 64-QAM usage limited to about 3\% for NR and 6\% for LTE.

% This is also seen in the uplink modulation usage for different technologies indoors and outdoors across all Sites-A,B,C, shown in Fig.~\ref{fig_phyUlModMnoNh}, where neutral-host has significantly higher usage of higher modulation (64-QAM) for 56\% of time, whereas for MNO macro indoors, it is 3\% for NR and 6\% for LTE.

\begin{figure}
    \centering
    \includegraphics[width=0.6\linewidth]{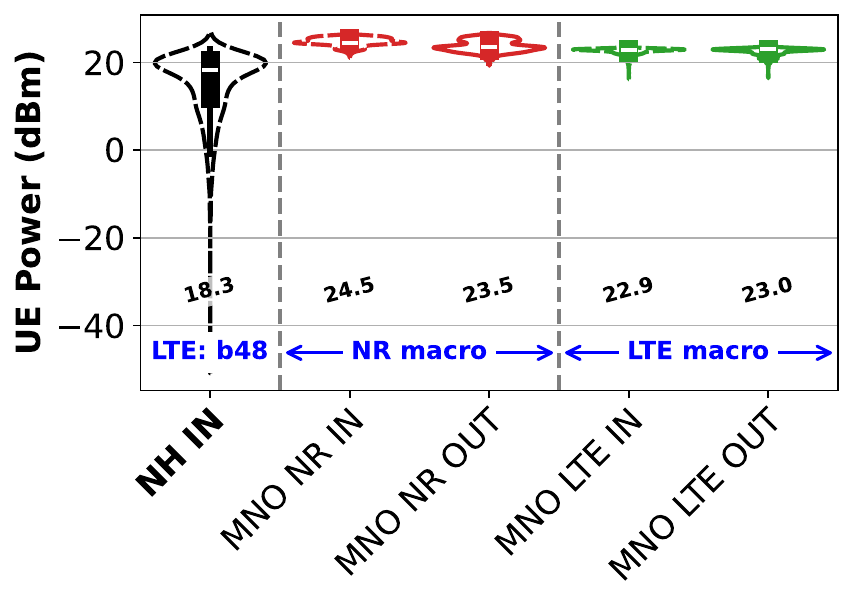}
    \caption{Uplink TX power.}
    \label{fig_phyUlTxMnoNh}
\end{figure}

A further advantage of the neutral-host deployment is the significant reduction in UE power consumption, as illustrated in Fig.~\ref{fig_phyUlTxMnoNh}. The NH UE utilized a median TX power of 18.3~dBm for uplink transmissions---approximately 5~dB lower than the median power required for MNO macro deployments. This improved power efficiency is primarily driven by the proximity of the UE to the indoor CBSDs, which minimizes path loss and allows the UE to maintain a robust link without reaching maximum transmit levels.

\begin{figure}
    \centering
    \includegraphics[width=0.6\linewidth]{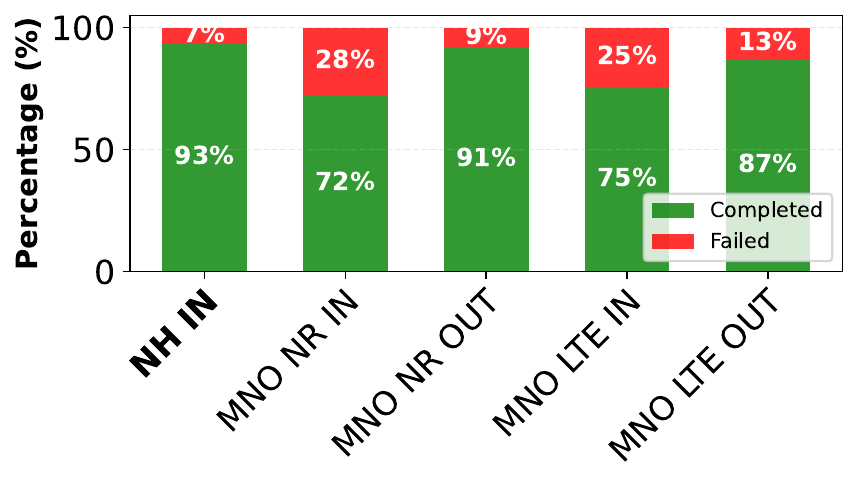}
    \caption{Uplink HTTP upload test status statistics.}
    \label{fig_phyPuschTestStatus}
\end{figure}

\begin{figure}
    \centering
    \includegraphics[width=0.6\linewidth]{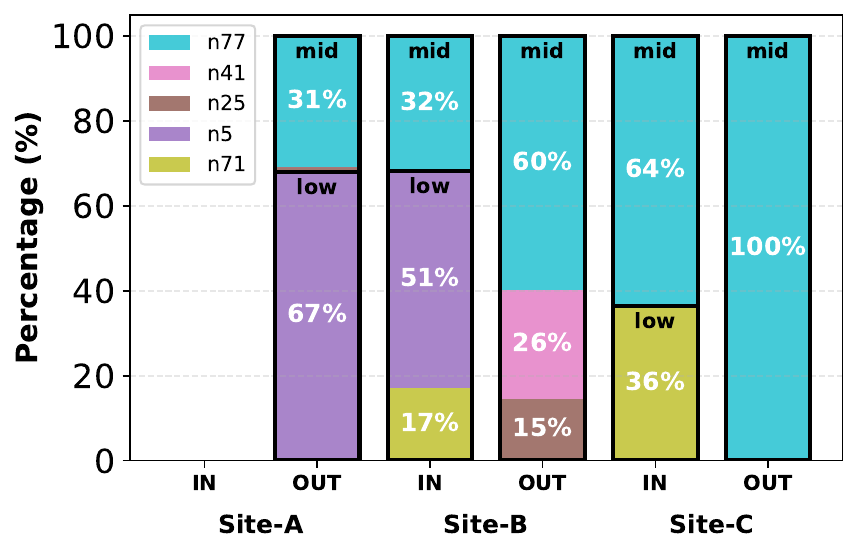}
    \caption{NR uplink band usage distribution across indoor and outdoor.}
    \label{fig_phyPuschBandUsage}
\end{figure}

Fig.~\ref{fig_phyPuschTestStatus} summarizes uplink test outcomes across all sites for neutral-host, MNO NR and LTE deployments. Results are based on QualiPoc active probe measurement tests consisting of HTTP upload to \textit{httpbin.org}, as described in Section~\ref{sec_methodology}. A test is classified as failed when the UE is unable to successfully establish a TCP connection with the target server. The analysis reveals that while outdoor MNO macro links experience relatively few failures, the failure rate increases indoors to approximately 28\% for NR and 25\% for LTE. In contrast, the neutral-host deployment achieves a markedly lower failure rate of about 7\%, closely matching the reliability observed for outdoor MNO macro scenarios. This highlights the robustness of the neutral-host uplink, particularly in indoor environments where macro coverage is more severely impacted.

\subsubsection{Band-Specific Analysis in the Uplink}\label{sec_uplink_bands}

To further investigate the role of frequency on uplink performance, we examine NR operations across low- and mid-band spectrum. The wider bandwidth of mid-bands (\textit{i.e.}, n41 and n77) enables superior peak rates. However, their indoor uplink remains constrained by the previously discussed propagation and power limitations, hindering their overall efficacy in indoor settings.

% The impact of low-band and mid-band operation on uplink performance in real-world deployments is examined by contrasting indoor and outdoor behavior. Focusing on NR uplink operation, 
Fig.~\ref{fig_phyPuschBandUsage} illustrates the relative usage of low- and mid-band spectrum across environments. For Sites-B,C, a pronounced shift toward mid-band utilization is observed in outdoor scenarios, with mid-band usage increasing from approximately 68\% indoors to complete usage outdoors at Site-B, and from about 36\% indoors to complete usage outdoors at Site-C. These observations indicate that, even when mid-band NR coverage is available---as evidenced by its dominant usage outdoors---UEs tend to avoid mid-band operation indoors. This behavior underscores the importance of low-band spectrum for indoor uplink, likely due to its superior propagation and penetration characteristics. At Site-A, no NR band usage is observed indoors, suggesting a comparatively weak NR deployment by the MNOs. Outdoors, NR operation at this site is dominated by low-band usage, indicating that, under limited deployment conditions, UEs preferentially select low-band spectrum to maintain reliable uplink connectivity.

\subsection{Comparison of End-to-End Network Performance}\label{sec_app_reports}

% \begin{figure}
%     \centering
%     \includegraphics[width=0.6\linewidth]{_figs/resultsv2/appTpRttNhWifiIn.pdf}
%     \caption{Application throughput and ping RTT analysis.}
%     \label{fig_appTpRttNhWifiIn}
% \end{figure}

\begin{figure}
    \centering
    \includegraphics[width=0.8\linewidth]{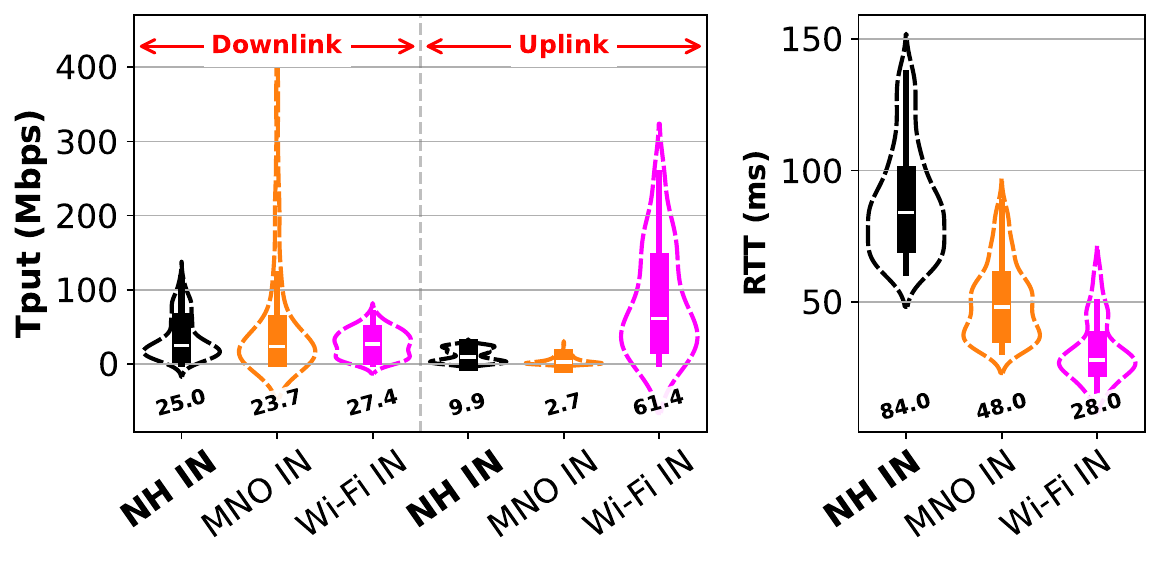}
    \caption{Comparison of network layer throughput and RTT.}
    \label{fig_appTpRttNhWifiMnoIn}
\end{figure}

To evaluate end-user performance and facilitate a direct comparison with Wi-Fi, we utilized application-based metrics using traffic described in \S\ref{sec_methodology}. Specifically, the metrics are measured at the network/IP layer, with DL and UL throughput measured using HTTP GET and POST traffic, respectively, while the round-trip time is measured via ICMP echo requests. Unlike PHY-layer metrics, which characterize raw radio capacity, these measurements capture end-to-end performance as observed by user traffic, including protocol overheads, scheduling effects, and network-side processing.

% IP throughput is measured using controlled HTTP GET and PUT transactions, corresponding to downlink and uplink traffic, respectively, while latency is assessed via ICMP ping measurements. Since these metrics are collected above the radio interface, the reported throughput reflects aggregate end-to-end performance rather than isolated link efficiency.

Fig.~\ref{fig_appTpRttNhWifiMnoIn} summarizes the distributions of indoor DL and UL throughput and round-trip time for NH, MNO macro, and enterprise Wi-Fi networks across all sites. The results indicate that NH median DL and UL throughput slightly surpass MNO macro levels, highlighting the efficacy of the NH model in overcoming indoor penetration loss. Nevertheless, Wi-Fi maintains a significant lead in UL throughput. On latency, there is a clear hierarchy with NH exhibiting the longest RTT, followed by MNO and Wi-Fi. This latency penalty in NH networks is primarily attributed to the additional routing overhead incurred during packet delivery to the MNO core.

\begin{table}
\centering
\captionsetup{font=small}
\caption{Site-wise indoor application-layer (IP) performance summary.}
\label{tab_appNhWifiSiteSummary}
\begin{tabular}{l l c c c c}
\toprule
\textbf{\makecell{Site}} &
\textbf{\makecell{Network}} &
\textbf{\makecell{\#PCIs\footnotemark[1] /\\ \#BSSIDs\footnotemark[2]}} &
% \textbf{\makecell{Backhaul}} &
\textbf{\makecell{DL Tput\footnotemark[3]}} &
\textbf{\makecell{UL Tput\footnotemark[3]}} &
\textbf{\makecell{RTT\footnotemark[3]}} \\
\midrule\midrule

\multirow{2}{*}{\textbf{Site-A}}
& NH   & 10 & 79.72 & 24.15 & 120 \\
& Wi-Fi& 67 & 34.9 & 44.02 & 46 \\
\addlinespace

\multirow{2}{*}{\textbf{Site-B}}
& NH   & 6 & 32.02 & 15.01 & 90 \\
& Wi-Fi& 244 & 44.54 & 161.02 & 28 \\
\addlinespace

\multirow{2}{*}{\textbf{Site-C}}
& NH   & 8 & 17.75 & 5.44 & 74 \\
& Wi-Fi& 230 & 19.46 & 24.56 & 25 \\
\bottomrule
\end{tabular}
\footnotetext[1]{Each CBSD is configured with two PCIs.}
\footnotetext[2]{2.4, 5~GHz counted separately.}
\footnotetext[3]{Median value.}
% \footnotetext[4]{The backhaul at the refinery location were shared between Sites-B,C and between Wi-Fi and NH.}
\end{table}

To further examine performance differences, a per-site breakdown between NH and Wi-Fi is provided in Table~\ref{tab_appNhWifiSiteSummary}. This comparison is particularly robust as both networks utilized identical backhaul infrastructure at each site and operated with comparable channel bandwidths.

% \todo{Write about backhaul details.}

% Across all three sites, the neutral-host deployment operates with carrier aggregation in the downlink for the overwhelming majority of measurements, yielding an effective downlink bandwidth of 40~MHz, which we treat as representative. In the uplink, neutral-host consistently operates with a 20~MHz bandwidth. In contrast, Wi-Fi operates with a 20~MHz channel at Site-A for both downlink and uplink, while at Sites-B,C, connected Wi-Fi measurements consistently use 40~MHz channels in both directions.

% These bandwidth configurations largely explain the observed DL throughput trends.
At Site-A, where NH operates at 40~MHz and Wi-Fi at 20~MHz for downlink, the neutral-host achieves an approximately \(2.3\times\) higher median DL throughput. At Sites-B,C, both technologies operate with comparable 40~MHz downlink bandwidth, with NH achieving a slightly lower median DL throughput (\(0.7\times\) at Site-B and \(0.9\times\) at Site-C). These trends underscore the primary influence of spectral allocation on downlink performance and demonstrate that NH can match enterprise Wi-Fi capacity with a significantly smaller hardware footprint, as evidenced by the lower CBSD count.

In the uplink, the trends diverge. At Site-A, both neutral-host and Wi-Fi operate with matching 20~MHz uplink bandwidth, yet Wi-Fi achieves an approximately \(1.8\times\) higher median UL throughput. This gap is further amplified in Sites-B,C, where NH operate at 20~MHz while Wi-Fi at 40~MHz. In these sites, Wi-Fi median UL throughput exceeds that of NH by factors of \(10.7\times\) and \(4.5\times\), respectively.
% This uplink advantage of Wi-Fi has also been reported in~\cite{palathinkal2025indoor}, which was conducted in a deployment setting closely matching Site-A---a large big-box retail environment---where both the Wi-Fi and NH deployments are effectively identical in terms of infrastructure and configuration.
% At Sites-B,C, Wi-Fi operates with double the uplink bandwidth (40~MHz) compared to neutral-host (20~MHz), but the observed UL throughput gains are substantially larger---\(10.7\times\) and \(4.5\times\), respectively---indicating that bandwidth alone does not fully account for the performance gap. 
A similar pattern is observed for round-trip time, where Wi-Fi consistently achieves lower latency across all sites.

Several factors likely contribute to the suboptimal UL throughput and latency performances observed in the NH network. These includes fundamental protocol differences: the 4G-based NH operates under TDD configuration that restrict uplink transmission to specific subframes, whereas Wi-Fi utilizes a contention-based mechanism. Additionally, neutral-host traffic must be routed to the respective MNO core network, whereas Wi-Fi traffic follows a more direct local path. These factors collectively constrain the NH uplink compared to the leaner Wi-Fi architecture.

%% file: _sections/conclusions.tex
\section{Conclusions and Future Works}

Our analysis of NH, MNO macro, and Wi-Fi deployments at there sites with distinct signal environments shows:

\begin{itemize}

% \noindent $\bullet$
\item There is a significant indoor coverage gap for MNO macro networks, with building penetration losses reaching 15.5~dB in the low-bands 17.9~dB in the mid-bands. With median mid-band signal RSRP trailing low-bands by roughly 10~dB in both indoor and outdoor environments, it is evident that low-band frequencies serve as the primary choice for indoor connectivity.
% , compensating for the propagation challenges faced by higher-capacity mid-band deployments.

\item Small-cell analysis at Site-A confirms that proximity drastically improves indoor coverage, with median RSRP gains of 34.5~dB and 38.1~dB for low- and mid-bands, respectively. Despite these gains, mid-band RSRP continues to trail low-band by up to 7~dB, mirroring the frequency-dependent propagation characteristics observed in macro deployments.

\item Per-PCI analysis further confirms the frequency-dependent nature of indoor attenuation by isolating signals from a single base-station source. Across both NR and LTE deployments, higher-frequency bands consistently exhibit larger indoor signal degradation than lower-frequency bands.

% \noindent $\bullet$
\item Due to a 30.3~dB improvement in median indoor RSRP over MNO macro-cells, the NH network delivers normalized indoor downlink throughput that exceeds indoor MNO benchmarks and matches MNO outdoor performance. The advantage is even more pronounced in the uplink, where the indoor NH solution surpasses MNO performance regardless of the MNO UE's environment (indoor or outdoor). 

% \noindent $\bullet$
\item NH networks utilize 64-QAM modulation for 56\% of indoor uplink transmissions, whereas MNO macro networks achieve this only 3\% (NR) and 6\% (LTE) of the time. Furthermore, NH UEs benefit from a 5~dB reduction in median transmit (TX) power. These performance gains are directly enabled by the proximity of UEs to the indoor CBSDs, which minimizes path loss and ensures a robust link budget.

\item While mid-bands are utilized extensively for outdoor 5G traffic, MNOs revert to low-band spectrum for the indoor uplink. This shift underscores the propagation constraints of mid-band frequencies, which often fail to maintain a reliable indoor uplink link budget due to high building attenuation.

% \noindent $\bullet$
\item The neutral-host network outperforms MNOs in indoor end-to-end throughput, but trails  Wi-Fi in UL throughput and latency. Specifically, Wi-Fi surpasses NH in median uplink throughput by up to \(1.8\times\) under matched 20~MHz bandwidth, and by up to \(10.7\times\) when utilizing \(2\times\) the NH bandwidth. These results underscore the necessity of protocol and routing optimizations to achieve full parity in uplink and latency-sensitive performance.

\end{itemize}

Looking ahead, a key area for further investigation is a deeper, controlled comparison between Wi-Fi and neutral-host systems. The results presented here are derived from customer-facing infrastructure, where deployment parameters cannot be systematically controlled to enable rigorous, apples-to-apples evaluation. To address this limitation, the University of Notre Dame has recently deployed a co-located experimental testbed comprising Wi-Fi~6E (operating in the unlicensed 6~GHz band) and neutral-host infrastructure (operating in the CBRS 3.5~GHz shared band). This testbed will enable targeted studies of several open research questions, including (i) the impact of neutral-host handover mechanisms---from MNO networks and across CBSDs---relative to inter-AP Wi-Fi handovers, and (ii) uplink throughput and latency behavior, which remain critical yet comparatively underexplored dimensions in Wi-Fi versus neutral-host performance evaluations.

\section*{Declarations}

\textbf{Author Contributions.}
The contributions are as follows: J.R.P. wrote the main manuscript text, performed the primary data analysis, and prepared all figures except for Fig. 2. M.I.R. contributed to the formal analysis and provided major text refinements and critical revisions. V.S. curated the source data and prepared Fig. 2. M.G. and M.Y. provided overall project supervision and secured the necessary resources. All authors reviewed and approved the final manuscript.

\vspace{.5em}\noindent
\textbf{Funding.}
Works in this paper by authors J.R.P., M.I.R., and M.G. are supported by the National Science Foundation (NSF) grant number CNS-2229387 and AST-2132700.

\vspace{.5em}\noindent
\textbf{Data Availability.}
The data supporting the findings of this study are available under license from Celona, Inc. and are not publicly available due to proprietary restrictions. However, anonymized datasets (stripped of specific operator and geospatial identifiers) are available from the corresponding author upon reasonable request and with the express permission of Celona, Inc.

\vspace{.5em}\noindent
\textbf{Competing Interests.}
The authors declare no competing interests.